\documentclass[sigconf]{acmart}
\AtBeginDocument{%
  }

\copyrightyear{2026}
\acmYear{2026}
\setcopyright{cc}
\setcctype{by}
\acmConference[CHI EA '26]{Extended Abstracts of the 2026 CHI Conference on Human Factors in Computing Systems}{April 13--17, 2026}{Barcelona, Spain}
\acmBooktitle{Extended Abstracts of the 2026 CHI Conference on Human Factors in Computing Systems (CHI EA '26), April 13--17, 2026, Barcelona, Spain}
\acmDOI{10.1145/3772363.3798729}
\acmISBN{979-8-4007-2281-3/2026/04}

\begin{document}

\title{Linguistic Similarity Within Centralized FLOSS Development}
\author{Matthew Gaughan}
\email{gaughan@u.northwestern.edu}
\affiliation{%
  \institution{Northwestern University}
  \city{Evanston}
  \state{Illinois}
  \country{USA}
}

\author{Aaron Shaw}
\authornote{Authors contributed equally to this work.}
\email{aaronshaw@northwestern.edu}
\affiliation{%
  \institution{Northwestern University}
  \city{Evanston}
  \state{Illinois}
  \country{USA}
}

\author{Darren Gergle}
\authornotemark[1]
\email{dgergle@northwestern.edu}
\affiliation{%
  \institution{Northwestern University}
  \city{Evanston}
  \state{Illinois}
  \country{USA}
}

\renewcommand{\shortauthors}{Gaughan et al.}

\begin{abstract}
When free/libre and open source software (FLOSS) stewards centralize project development, they potentially undermine project sustainability and impact how contributors talk to each other. To study the relationship between steward-centralized development and contributor discussion, we compared the development of three Wikimedia platform features that the Wikimedia Foundation (WMF) built in MediaWiki. In a mixed-methods multi-case comparison, we used repository mining, linguistic style features, and principal component analysis to track MediaWiki feature development and issue discussions. Contrary to both our intuition and prior work, there were no identifiable differences in the linguistic style of WMF-affiliates and external contributors, even when feature development was  guided by WMF contributions. From these results, we offer two provocations to the study of collaborative FLOSS development: (1) stewards dominate development according to their own use of specific project functionality; (2) centralized project development does not entail hierarchical language within project discussions. 
\end{abstract}





 \maketitle

\section{Introduction}
Though free/libre and open source software (FLOSS) stewards assume primary responsibility for project development, too many steward contributions may risk the very sustainability that such development activity is meant to support. Increasingly common among large FLOSS projects \cite{germonprez_eight_2018, osborne_open_2024, li_systematic_2025}, FLOSS \textit{stewardship} describes large nonprofits or for-profit firms which take responsibility for the maintenance, development, and distribution of a focal FLOSS project. Stewardship can centralize project decisions among steward-affiliated contributors, leading to \textit{steward dominance} of project development work \cite{osborne_characterising_2025}. As with any FLOSS project, such centralized development can stratify internal communication, marginalize peripheral voices, and lead to community decay \cite{han_characterizing_2024, barcomb_managing_2022, zhou_how_2020}.
Consider Oracle's stewardship of OpenOffice: in response to Oracle's tight control of OpenOffice's software development, external contributors eventually forked the project to create LibreOffice \cite{gamalielsson_sustainability_2014}. By dominating project engineering, Oracle undermined its own stewardship: following the LibreOffice fork, Oracle donated the OpenOffice project to the Apache Software Foundation. More broadly, the productive capacity towards a free/open source alternative to Microsoft Office was fractured by the fork and remains split today.

To study FLOSS stewardship and its relationship to collaborative communication within a project, we compared the development of three Wikimedia platform features which were built by the Wikimedia Foundation (WMF) in MediaWiki, the open source wiki framework. We analyzed feature deployment announcements, MediaWiki commit histories, and conversations in the project issue tracking system. Modeling the linguistic style of contributor discussions, we measured comments' stylistic features and analyzed conversational registers with principal component analysis.

We found that WMF-affiliates authored the overwhelming majority of feature-relevant contributions even as external contributors authored most of the commits to MediaWiki as a whole. Contrary to both our intuition and prior work, there were no identifiable differences in the linguistic style of WMF-affiliates and external contributors in discussions during this WMF-centered development; even as WMF-affiliates made decisions in off-platform, internal meetings.  
From these results, we offer two contributions to the study of collaboration within FLOSS development. First, we illustrate that steward contribution activity may be largely contingent on stewards' own-use motivations of specific project functionality. Second, we propose that centralized project development does not entail stratified linguistic registers between project contributors. 

\section{Related Work}
Though there are a few different models of FLOSS stewardship, the steward organization often assumes a central role within project software engineering. In single-vendor stewardship, the steward develops and packages the focal FLOSS project while simultaneously producing a downstream implementation of it \cite{riehle_open_2021, zhang_companies_2021, li_systematic_2025, riehle_single-vendor_2020}. As both the primary developer and a major user of the library, single-vendor stewards often dominate the authorship of project commits and centralize technical decision-making as they develop the library towards their own use \cite{osborne_characterising_2025, orucevic-alagic_network_2014, teixeira_collaboration_2014, zhang_chromium_2023}. This can be an issue for external contributors who are not affiliated with the steward organization. Though the upstream project may be open and accessible to all, steward-affiliates often make use of internal meetings away from accessible platforms, obfuscating project decisions from external contributors \cite{osborne_characterising_2025}. Although the use of off-platform meetings may disrupt public development conversations on the project issue tracking system (ITS) \cite{arya_analysis_2019}, steward dominance is often identified in the authorship of project commits \cite{zhang_companies_2021,osborne_characterising_2025}. Project collaboration is structured by rules defining who gets to deliberate decisions and who gets to make them. However, the full impact of formal voice and decision rights in FLOSS settings remains to be seen \cite{freeland_problems_2018, turco_conversational_2016}. Given single-vendor stewards' use of the upstream project in other software products, we propose that stewards will exercise a lot of voice and decision-making ability in project development. 

Prior work suggests that such centralized development should stratify contributor communication and negatively impact project sustainability. As development is centralized, peripheral contributors are marginalized. In some cases, marginalization can lead to contributors exiting the project entirely \cite{gamalielsson_sustainability_2014, zhou_how_2020}. This exclusion may also shape project development conversations.  In social linguistics, the intent of a statement and the speaker's social relationship to the interlocutor define conversational language into a given \textit{register} \cite{biber_register_2009}. Prior empirical studies of communication within FLOSS development suggests that differences in the registers of project discussions can indicate informal project hierarchies. Though one study found little difference in the linguistic style of elite and non-elite FLOSS contributors within ITSs \cite{han_cross-status_2023}, a more recent paper by the same authors suggests that project hierarchy correlates to distinct linguistic styles that are prevalent across semantic contexts \cite{han_characterizing_2024}. Moreover, in the code review process of the Linux Kernel, socially privileged contributors can omit otherwise important information in their patches and still succeed in change acceptance \cite{tan_how_2019}. As such, we expect that any steward-centralized development will result in steward-affiliated and external contributors communicating differently within project ITS discussions. 

\section{Study Design}
To study centralized development and contributor communication within a stewarded FLOSS project, we completed a multi-case comparison of three own-use feature deployments within a large stewarded FLOSS project. Across all three cases, we collected trace data on commits and issue tracking discussions, which we then analyze using mixed methods in order to compare the social dynamics of contributions and technical discussion.
\subsection{Case Selection}
\label{methods:case-selection}

We followed Eisenhardt's research design recommendation to focus on cases where the focal phenomena can be observed and are likely to occur \cite{eisenhardt_what_2021}. To study centralized FLOSS development, we followed the rationale that within a single-vendor FLOSS project, own-use feature development may be more steward-controlled than other work. As such, we selected three Wikimedia feature deployments built within MediaWiki. 

\href{https://www.mediawiki.org/wiki/MediaWiki}{MediaWiki} is a widely used FLOSS project used to build wiki websites such as Wikipedia. The Wikimedia Foundation (WMF) stewards and develops MediaWiki in the service of producing Wikimedia platforms.\footnote{\href{https://www.mediawiki.org/wiki/Manual:What_is_MediaWiki}{\textit{MediaWiki.org Manual: What is MediaWiki?}}} We selected three MediaWiki features deployed on Wikimedia: \textit{VisualEditor}, \textit{HTTPS-login}, and \textit{HTTP-deprecation}. \textit{VisualEditor} is an graphical user interface (GUI) for editing wikitext, the markup language for MediaWiki. Deployed to Wikipedia on July 1, 2013, The primary functionality of VisualEditor was developed within the \texttt{extensions/visualeditor} MediaWiki library. \textit{HTTPS-login} established HTTPS was the default communication protocol for all logged-in users on Wikimedia. Deployed to Wikipedia on August 28, 2013, this feature required changes to default user login functionality within the widely used \texttt{mediawiki/core} library. \textit{HTTP-deprecation} removed all remaining HTTP functionality with the deployment of HTTP Strict Transport Security (HSTS). Deployed on July 2, 2015, this feature also required changes to the default authentication of \texttt{mediawiki/core}. For all three cases, we defined the deployments as spanning from one month prior to the announcement of the feature's alpha opt-in testing (e.g. HTTPS' opt-in testing was announced on October 3, 2011, we collect data from September 3), to 90 days after the feature is deployed in a widespread opt-out form (e.g. VisualEditor was deployed on July 1, 2013, we collect data until September 29.)
\subsection{Data Collection}
\label{methods:data-collection}
We collected project development data through Wikimedia's primary ITS (Phabricator) and the \texttt{mediawiki/core} and \texttt{extensions/ VisualEditor} MediaWiki repositories. Our data and code are available on the Harvard Dataverse at \href{https://doi.org/10.7910/DVN/TKTVF9}{\textbf{this hyperlink}}.

\begin{figure*}[t!]
    \centering
        \includegraphics[width=\linewidth]{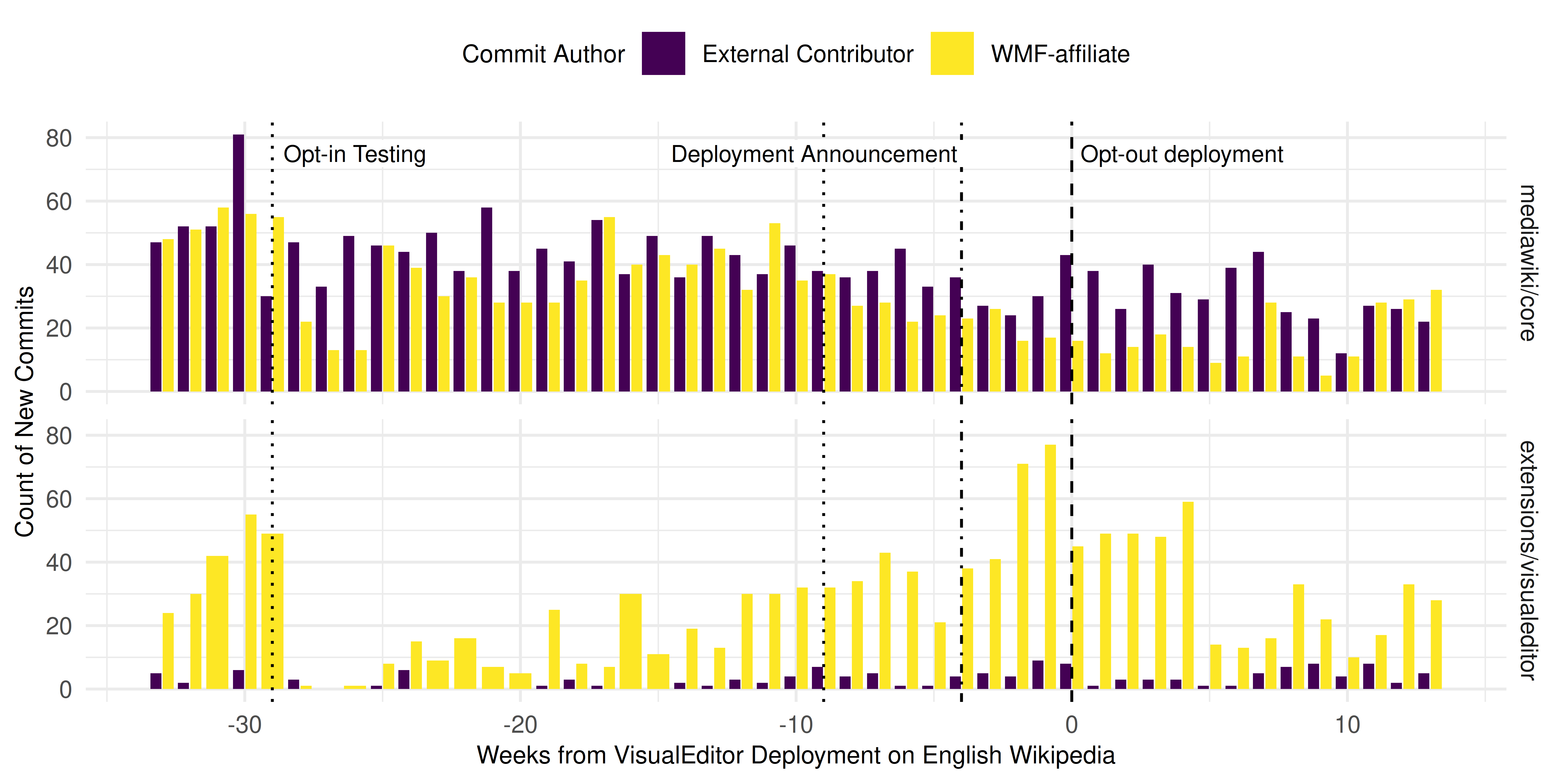}
        \caption{Weekly count of commits by contributor affiliation for the duration of VisualEditor's Wikimedia deployment. Faceted by repository, we can see that at the same time that WMF-affiliates author the vast majority of commits to \texttt{extensions/visualeditor}, they author the minority of commits to \texttt{mediawiki/core}.} 
        \label{fig:ve_commits_created}
        \Description[More WMF-authored commits in VisualEditor repository than in core repository during the same period.]{Bar plots of the weekly counts of commits by either WMF-affiliates or volunteers. WMF-affiliates author an overwhelming majority of commits to VisualEditor's repository while authoring the minority of commits to MediaWiki core during the same period.}
    \end{figure*}
\subsubsection{Phabricator Collection}
To identify relevant Phabricator Tasks for each feature, we used natural language key-term searches rather than user-edited labels. At the time of our feature deployments, these user-generated categorical labels were not always appropriately applied, especially when issues were migrated from Bugzilla.\footnote{Out of the 3126 issues in our final data set, 246 (7.9\%) of them were imported by the \texttt{bzimport}; our of the total 21901 task replies, 923 (4.2\%) of them were imported.} As such, our data sampling relied on key-term searches instead.\footnote{We initially queried for the simple key-terms ``\texttt{visualeditor}'' and ``\texttt{http*}'' during the timeframes of our deployment cases. Given the high amount of irrelevant tasks returned for ``\texttt{http*}'', we only included returned items with titles or descriptions that contained one of the following tokens: \texttt{http*}, \texttt{login}, \texttt{cert}, \texttt{ssl}, \texttt{tls}.} In total, we collected 3,126 relevant Tasks with 21,901 comments.

We identified Phabricator accounts' WMF affiliation in two moments of data collection. The first moment was a query for all Phabricator accounts that contained the substring ``WMF'' in their usernames on February 28, 2025; this resulted in a list of 324 accounts. The second moment of data collection consisted of a manual search through available team rosters, LinkedIn profiles, and engineering reports. In total, we identified 463 Phabricator accounts belonging to current or former WMF affiliates. However, not all of these affiliates contributed to Task discussions within our feature cases. Across the 741 Phabricator contributors across our three cases, we identified 109 WMF-affiliates.\footnote{VisualEditor was developed in collaboration with Wikia (now Fandom.) Given this formal organizational coordination was only in place for the VisualEditor case, we identify Wikia contributors as WMF-affiliates for the VisualEditor case but as external contributors in the other two deployments. Given the two organizations' close collaboration in the VisualEditor case, we argue that Wikia is a de facto affiliate of the WMF during VisualEditor's development.} 

\subsubsection{Commit Data}

We collected the full git histories for the \texttt{main} branches of two of the central MediaWiki library \texttt{mediawiki/core} and the secondary \texttt{extensions/visualeditor}. For commit data, we used email signatures to identify organizational affiliation of committers to the relevant code repositories. All authors with \texttt{@wikimedia.org} or \texttt{@wikimedia.de} emails were labeled as WMF-affiliates. Wikia contributors were identified by the \texttt{@wikia-inc.com} signature. In addition, we manually parsed through the repository histories and identified six WMF employees who were using personal email accounts. We present the commit histories of \texttt{core} and \texttt{visualeditor} during VisualEditor's deployment in Figure \ref{fig:ve_commits_created}.

\subsection{Modeling Communication}
\label{methods:modeling-communication}
\subsubsection{Linguistic Style}

Following prior work studying communication within FLOSS projects \cite{han_cross-status_2023,han_characterizing_2024}, we applied computational methods to identify linguistic style characteristics within ITS conversations. Pilot evaluations with simpler measures such as the use of hedging words found little difference between contributor populations. Subsequently, we used \texttt{BiberPlus}, a Python toolikit that measures 96 features derived from Biber's metrics of linguistic style \cite{biber_variation_1988, alkiek_neurobiber_2025}. We then fit principal components (PCs) to account for 90\% of the variance within our \texttt{BiberPlus} feature vectors \cite{pedregosa_scikit-learn_2011}; this resulted in 25 PCs. 

With 25 PCs, many of the PCs accounted for small amounts of stylistic variability. As such, we only manually interpreted the four PCs which accounted for the most variance: lengthy discussion to brief updates (PC1); a proxy measurement of sentence length (PC2); expressive, first person comments to dry, third person statements (PC3); and technical jargon to non-technical observations (PC4). 

\subsubsection{Conversational Registers}

To study ITS registers, we grouped comments by intent of the statement and the social position of the speaker \cite{biber_register_2009}. Within any ITS discussion, interlocutors may have a range of motivations and goals for their comments. First, we defined a set of comments which we reasoned shared similar functional orientation: all comments made by the Task author prior to the Task's resolution are likely in the service of the Task being resolved. Inclusive of the Task description and and all subsequent comments, we label this set of comments as \texttt{requests}.\footnote{If the Task is unresolved, then \texttt{requests} defines \textit{all} comments made by the Task Author.} Second, we compared comments made by WMF-affiliates and external contributors. The rationale for this comparison was that steward-affiliates and external contributors may talk differently to each other if affiliates are dominating project development.

\section{Results}

\begin{figure*}[t!]
    \centering
        \includegraphics[width=\linewidth]{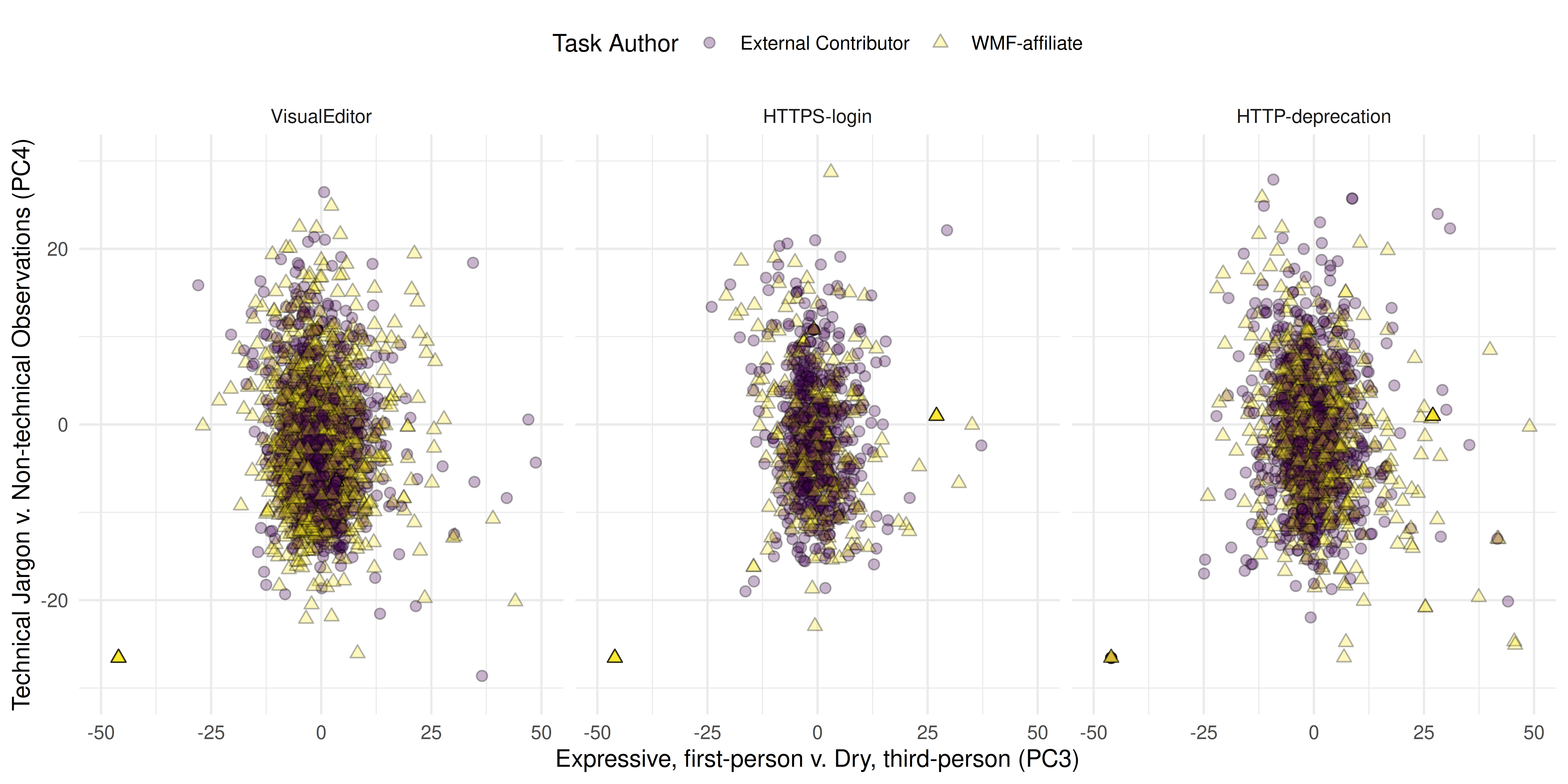}
        \caption{Affiliation-attributed Phabricator comments (\texttt{requests}) plotted along PC3 and PC4 of a PCA on \texttt{BiberPlus} features. All comments are clustered similarly, indicating no difference between authors' organizational affiliation and linguistic register.} 
        \Description[No distinction between volunteer and WMF-affiliate PC clusters]{Plotted within a two-dimensional PC3 and PC4 space, there is no identifiable difference in the clustering of comments from WMF-affiliates or volunteer contributors.}
        \label{fig:request_pc}
\end{figure*}

Across all three of our feature cases, we found that the WMF assumed a primary role in feature development, more so than in the development of other MediaWiki functionality during the same time period. Our manual search identified at least twelve changes to the default functionality of \texttt{mediawiki/core} in support of case features' deployment. The default MediaWiki interface, IP-variable logins, and URL handling were all changed in explicitly own-use commits. Almost all of the commits to \texttt{mediawiki/core} that we identified as being motivated by a focal feature were authored by WMF-affiliates; one contribution during VisualEditor was made by a volunteer who would later be employed by the WMF. This WMF dominance contrasts from its role in broader MediaWiki development. As shown in Figure \ref{fig:ve_commits_created}, the WMF authored over 90\% of commits to \texttt{extensions/visualeditor} while authoring a minority of the commits to \texttt{mediawiki/core} during the same time period. Broadly, out of the 11,712 non-bot commits made to \texttt{core} between between November 11, 2012 (30 days prior to VisualEditor's opt-in testing announcement) to September 30, 2015 (90 days after HTTP-deprecation was deployed) WMF-affiliates authored 4,915 (42.0\%) commits, Wikia contributors authored two commits, and external contributors authored 6,795 (58.0\%) commits.   

Though the development of Wikimedia features was more WMF-dominated than other MediaWiki development, both within-case and between-case comparisons of features' ITS discussions identified no stylistic difference between contributor populations. We visualize this in Figure \ref{fig:request_pc}. Across our four-dimensional principal component space (PC1, PC2, PC3, PC4,) the Euclidean distance between the WMF-affiliate and external contributor group \texttt{requests} medoids is 0.47 while the mean absolute deviation across the PCs is 6.24; this separation-to-spread ratio of 0.08 indicates little difference between contributor population clusters. Further, a one-way MANOVA revealed no significant effect of affiliation on the four-dimensional PCA space ($Pillai=0.000$; $p=0.28$). This between-population similarity held when we compared affiliation clusters across Task descriptions and replies as well. 

This finding is of interest given that WMF employees sometimes referred to offline, closed-door meetings in which technical choices such as feature prioritization or issue refusal were decided. Any ITS comments referencing these meetings were for posterity. Comments such as: ``\textit{we decided in the meeting to remove [this Task] from the Q3 blockers list},'' ``\textit{Short term solution as decided by [WMF employee], [Wikia employee] and [WMF employee]: Make links not auto-extend in the model,}'' and ``\textit{This was discussed at the weekly meeting on 2015-06-30. We decided that it wasn't a priority for this quarter.}'' suggested that Phabricator served a different function for WMF-affiliate contributors than it did for external contributors. If, as a function of meaningful deliberation in these off-Phabricator meetings, WMF-affiliates primarily commented brief updates or post-hoc explanations, we would see well-defined contributor affiliation clusters within our stylistic PCs. As shown in Figure \ref{fig:request_pc}, we see no such clustering. Rather, although we have identified WMF-affiliates' use of off-platform communication and dominance within these features' development, we see no difference between the style of WMF-affiliate's issue requests and those of external contributors.



\section{Discussion}

The prevalence of WMF-authored commits within our feature case studies aligns with prior literature, which has found that stewardship organizations often assume the central role in the development of their focal FLOSS project \cite{osborne_characterising_2025, zhang_companies_2021}. Advancing prior work, we found that this steward development is not uniform across project functionality in single-vendor stewardship. Instead, WMF-affiliates authored a larger share of feature-relevant commits than they did for project development more broadly, suggesting that steward dominance may be related to own-use relevance of specific project features. More research is necessary to map steward attention across project functionality and identify any downstream risk from irregular development: varying domains of steward attention may result in similarly variable project maintenance and software security.

In conversations regarding features where WMF-affiliates accounted for the majority of contributions, we found no stylistic difference between WMF-affiliates and external contributors on feature-relevant Phabricator discussions. On one hand, this between-population similarity is surprising given Han et al.'s \cite{han_characterizing_2024} recent paper which identified that the communication styles of socially advantaged and disadvantaged FLOSS contributors are different. On the other hand, Han et al. suggest that firm stewardship may limit between-population communication differences as structured work processes counteract projects' social hierarchies. 

We offer a different hypothesis for the lack of linguistic stratification in contributor discussions. The WMF work processes that we observed centralized feature development while simultaneously decentralizing project communication. The WMF authored the majority of commits, controlled high-level technical decisions,\footnote{\href{https://www.mediawiki.org/wiki/Wikimedia_Engineering/2014-15_Goals}{2014-2015 Wikimedia Engineering Goals}} and controlled Phabricator Task prioritization.\footnote{\href{https://www.mediawiki.org/w/index.php?title=Bugwrangler&oldid=868269}{MediaWiki: Bugwrangler (January 2014)}} Affiliate contributors accessed privileged project decisions in ways that external contributors could not. Nevertheless, like other FLOSS stewards that used off-platform meetings \cite{osborne_characterising_2025}, the WMF seems to have kept the majority of MediaWiki communication on the project ITS.\footnote{\href{https://www.mediawiki.org/w/index.php?title=Phabricator/Project_management&direction=next&oldid=1304622}{MediaWiki:Phabricator: Project Management (December 2014)}} Though the WMF dominated feature decisions and contributions, many external contributors were able to voice their perspectives through Phabricator. We propose that this model exemplefies Turco's proposal of decoupled voice and decision rights, where “voice and dialogue are open but decision making is not”\cite{turco_conversational_2016}.\footnote{Future research is necessary to evaluate the sustainability of such a model: in the years following our case studies, the WMF has attempted to \href{https://www.mediawiki.org/wiki/Technical_decision_making}{decentralize project decision making}.} 


\section{Limitations}

With all three of our feature deployment cases occurring within the Wikimedia-MediaWiki project, the generalizeability of our findings are limited by our empirical setting. However, this limitation is inherent to any theoretically-sampled case study. Moreover, the Wikimedia ecosystem has a substantial history as a setting for academic research of governance and collaboration within online peer-production \cite{hwang_rules_2022, keegan_hot_2011}.

The timing of our three cases risks constraining our findings to a specific era in FLOSS development. While all of our cases take place between Autumn 2012 and Autumn 2015, we do not believe that this similarity limits our findings' validity. Though FLOSS norms and governance have changed since 2015, our findings of steward dominance and mitigated communication differences align with prior empirical work studying more recent FLOSS settings \cite{osborne_characterising_2025, han_characterizing_2024}.

\section{Conclusion}
Through our case studies of MediaWiki feature developments, we make two primary contributions to the study of collaborative FLOSS development. First, we identify own-use feature development as \textit{more} steward-dominated than general library contributions, suggesting that within a project's functionality, there are different domains of steward development attention. Second, we propose that centralized development does not entail stratified project linguistic registers. 

\begin{acks}
    We thank the members of the MediaWiki and Wikimedia communities who have made their data available to the public. We gratefully acknowledge the support of a grant from the Digital Infrastructure Insights Fund, fiscally supported by Aspiration. We would also like to thank Rob Voigt for his aid in this study's linguistic analysis and Martin Gerlach for his support on an earlier version of this work. This work was conducted using the Hyak supercomputer at the University of Washington as well as research computing resources at Northwestern University.
\end{acks}


\bibliographystyle{ACM-Reference-Format}
\bibliography{chiea-cites.bib}


\end{document}